\documentclass[]{spie}  
\usepackage{graphicx}
\usepackage{rotating}
\usepackage{amssymb}
\usepackage{mathptmx}
\usepackage{epstopdf}
\usepackage[]{natbib}
\usepackage{lineno}
\usepackage{url}

\bibpunct{[}{]}{,}{n}{}{,}

\begin{document}

\newcommand{\hdblarrow}{H\makebox[0.9ex][l]{$\downdownarrows$}-}
\newcommand{\pb}{\protect\textsc{polarbear}}
\newcommand{\Pb}{\protect\textsc{Polarbear}}

\title{Commercialization of micro-fabrication of antenna-coupled Transition Edge Sensor bolometer detectors for studies of the Cosmic Microwave Background}

\author{   
Aritoki Suzuki$^{a*}$,
Chris Bebek$^{a}$, 
Maurice Garcia-Sciveres$^{a}$, 
Stephen Holland$^{a}$, 
Akito Kusaka$^{a,e}$, 
Adrian T. Lee$^{a,b,c}$, 
Nicholas Palaio$^{d}$, 
Natalie Roe$^{a}$, 
Leo Steinmetz$^{b}$
\skiplinehalf
\small{
$^{a}$ Physics Division, Lawrence Berkeley National Laboratory, Berkeley, CA 94720, USA
\\$^{b}$ Department of Physics, University of California, Berkeley, CA 94720, USA
\\$^{c}$ Radio Astronomy Laboratory, University of California, Berkeley, CA 94720, USA
\\$^{d}$ Engineering Division, Lawrence Berkeley National Laboratory, Berkeley, CA 94720, USA
\\$^{e}$ Department of Physics, University of Tokyo, Tokyo 113-0033, Japan
\\$^{*}$ Corresponding author: asuzuki@lbl.gov 
}
}


\maketitle

\begin{abstract}
We report on the development of commercially fabricated multi-chroic antenna coupled Transition Edge Sensor (TES) bolometer arrays for Cosmic Microwave Background (CMB) polarimetry experiments. 
CMB polarimetry experiments have deployed instruments in stages.
Stage-II experiments deployed with O(1,000) detectors and reported successful detection of B-mode (divergent free) polarization pattern in the CMB. 
Stage-III experiments have recently started observing with O(10,000) detectors with wider frequency coverage.
A concept for a Stage-IV experiment, CMB-S4, is emerging to make a definitive measurement of CMB polarization from the ground with O(400,000) detectors.
The orders of magnitude increase in detector count for CMB-S4 requires a new approach in detector fabrication to increase fabrication throughput.and reduce cost. 
We report on collaborative efforts with two commercial micro-fabrication foundries to fabricate antenna coupled TES bolometer detectors.
The detector design is based on the sinuous antenna coupled dichroic detector from the POLARBEAR-2 experiment. 
The TES bolometers showed the expected I-V response and the RF performance agrees with simulation.
We will discuss the motivation, design consideration, fabrication processes, test results, and how industrial detector fabrication could be a path to fabricate hundreds of detector wafers for future CMB polarimetry experiments.
\keywords{Cosmic Microwave Background, TES bolometer, Micro fabrication, Inflation, Polarization, B-mode}
\end{abstract}

\section{Introduction}
Over the past two decades, teams from around the world have made increasingly sensitive measurements of the Cosmic Microwave Background (CMB) using telescopes on the ground, balloons and satellites. 
Its uniformity confirmed the hot big bang model of the universe, and detailed measurements of its small $10^{-5}$ non-uniformity have led to tight constraints on the composition, geometry, and evolution of the universe. 
The CMB is also weakly polarized through Thomson scattering by free electrons, and it provides access to information that is inaccessible from temperature anisotropy \cite{S4ScienceBook}.

CMB polarization can be decomposed into two orthogonal bases that are determined by the physical origins. 
Scalar perturbations that are also responsible for temperature anisotropy and structure formation produced the parity-conserving polarization pattern (E-mode). 
The existence of light particles such as neutrinos and postulated light relics, including axions and sterile neutrinos, alters the total energy density of the radiation in the radiation-dominated era of the universe to distort the E-mode power spectrum at arc-minute angular scales.
At very large angular scales, the E-mode is a probe to the physics of reionization that would allow CMB measurement to tighten constraint on the sum of neutrino masses. 
Starting in late 2013, ground-based CMB experiments reported the detection of the parity-violating polarization pattern in the CMB, called the B-mode. 
Large-scale structures between the surface of last scattering and the observer deflect the E-mode polarization pattern to convert E-mode into B-mode through weak gravitational lensing. 
This {\it lensing} B-mode peaks at an arc-minute angular scale, and amplitude of its angular powerspectrum constrain the sum of neutrino masses. 
At degree angular scale, amplitude of B-mode spectrum will allow us to explore to models of inflation, a postulated time period of rapid expansion of the universe at early universe, if its energy scale was at detectable level. 
To explore these sciences and beyond, the field deployed CMB polarimetry experiments in stages.
Stage-II experiments deployed with O(1,000) detectors that reported successful detection of B-mode (divergent free) polarization pattern in the CMB. 
Stage-III experiments have recently started observing with O(10,000) detectors with wider frequency range coverage to characterize galactic foregrounds.
A concept for a Stage-IV experiment, CMB-S4, is emerging to make a definitive measurement of CMB polarization from the ground with O(400,000) detectors.

Concepts of configurations for the next generation CMB experiment have been studied by the CMB communities \cite{CDT, LBNLStudy}.
Studies found that O(400,000) detector spread across 20 GHz to 300 GHz on a few large aperture (6 meters) telescopes and over a dozen small aperture ($\approx$ 0.5 meter) telescopes deployed at Chile and South Pole sites, two sites that have current CMB ground based telescopes, is a configuration that can achieve the science target of the next generation CMB experiment. 
Concept Definitioin Task Force reported that this next generation CMB experimental configuration will achieve $\sigma(r) \leq 0.0005$, $\sigma(\Delta N_{eff}) < 0.03$, and $\sigma(\Sigma(m_\nu)) < 25~\mathrm{meV}$ with five to seven years of observations. 

To deploy O(400,000) detectors, the Stage-IV experiment will require approximately 300 150-mm yielded detector wafers. The actual number of wafers that will need to be fabricated will depend on production yield.
Up until now, CMB detector wafers were micro-fabricated in clean rooms at the national laboratories and the universities by specialized enigneers, technicians, postdocs, or graduate students.
The orders of magnitude increase in wafer count for the Stage-IV experiment requires a new approach in detector fabrication toincrease throughput. As a
potential new approach to fabricate CMB detectors,  we worked collaboratively with two commercial micro-fabrication foundries to fabricate antenna coupled TES bolometer detectors. 
Industrial fabrication facilities have a potential to increase throughput and reduce cost per wafer by having more stream lined processes and multiple shifts.
Also, stringent industrial approach on quality assurance could improve uniformity, repeatability, and yield that are necessary for the next generation CMB experiment.
We designed, fabricated, and tested prototype detectors with commercial foundries to demonstrate this new approach.

\section{Design and Fabrication}
\begin{figure}[!h]
\begin{center}
\includegraphics[width=\textwidth,keepaspectratio]{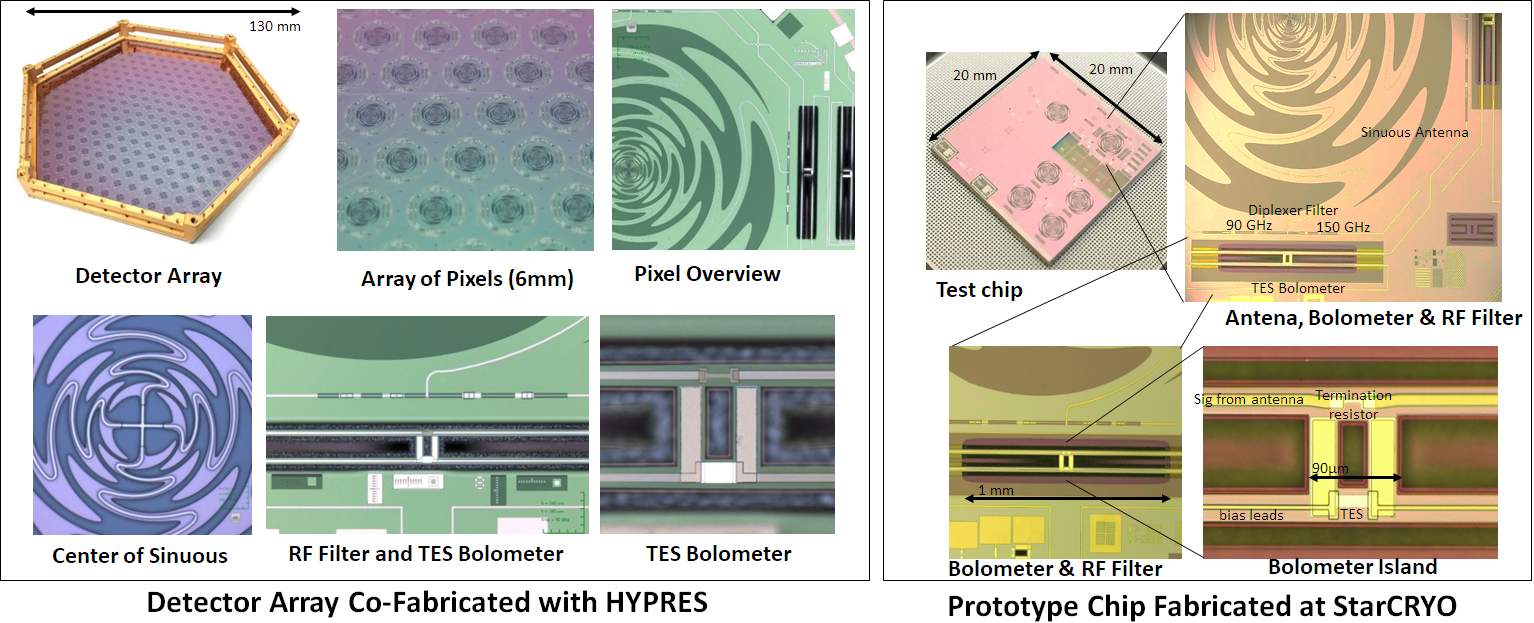}
\end{center}
\caption{(Color online) [Left] Photographs of sinuous antenna coupled detector array fabricated with HYPRES. Top row from left: hexagonal detector array mounted in POLARBEAR-2 detector holder,  zoomed in array of pixels that are 6.8 mm apart, and overview of a pixel with sinuous antenna and TES bolometer highlighted. Bottom row from left: zoom in photograph at center of sinuous antenna showing 2 micron wide feed, RF diplexer filter and TES bolometer, and zoom in at center of bolometer island. Bright white rectangle in the zoomed in photograph of the bolometer island is an AlMn TES bolometer. [Right] Photograph of sinuous antenna coupled detector test chip fabricated by STAR Cryoelectronics. Several test devices were fabricated on 20 mm by 20 mm chip. Zoomed in photographs show pixel overview, including the sinuous antenna, RF filter, and TES bolometer.}
\label{fig:waferphoto}
\end{figure}

Micro-fabrication of science grade detectors by commercial foundry was successfully implemented by the Dark Energy Spectroscopic Survey experiment where Lawrence Berkeley National Laboratory and Teldyne DALSA Inc. co-fabricated CCD devices \cite{DESI}. 
Modern micro-fabricated CMB detectors require superconducting niobium.
We collaborated with two commercial foundries that specialize in micro-fabrication of niobium devices. 
Two foundries, HYPRES Inc. and STAR Cryoelectronics, specialize in fabrication of superconducting devices such as Superconducting QUantum Interference Devices (SQUIDs) and voltage standards with Josephson junctions \cite{hypres,starcryo}. 
Both foundaries were familiar with films that are commonly used in the CMB detector fabrication such as niobium, silicon dioxide, aluminum, titanium, palladium, and silicon nitride. 
The detector design was based on the sinuous antenna coupled dichroic (90 GHz and 150 GHz) detector from the POLARBEAR-2a experiment \cite{SuzukiLTD15}.
20 mm by 20 mm test chips were fabricated at STAR Cryoelectronics and HYPRES Inc. co-fabricated 271 pixel detector arrays, as shown in Figure~\ref{fig:waferphoto}.

Detectors were fabricated on 150 mm diameter silicon wafers at both foundries. 
Fabrication steps were based on the POLARBEAR-2 detector design that was developed at the Marvell Nanofabrication Laboratory of the University of California, Berkeley \cite{WestbrookLTD17}. 
STAR Cryoelectronics processed every step of fabrication at their facility.
STAR Cryogelectronics devices were fabricated with aluminum Transition Edge Sensors (TESs). 
HYPRES co-fabricated wafers with the UC Berkeley Nanofabrication Laboratory.
Most processes were completed at HYPRES, then deposition of aluminum-manganese film and bolometer release step with Xenon difluoride gas were done at the UC Berkeley Nanofabrication Laboratory. 
Fabrication of each layer requires deposition, lithography, etch, clean, and characterization. 
Both fabrication foundries spent a day per layer with a single shift per day. 
We can further increase production rate by increasing the number of shifts per day to meet future demands. 
Absolute film thickness and its uniformity across wafers were measured with ellipsometer and profilometer during fabrication processes. 
The silicon oxide film has the tightest tolerance, as its variation in thickness across a wafer will result inshifts of the center frequency of
the detector’s band pass filter. 
Film thickness variation of this film was less than 2\% which meets the current experiment's specification.
STAR Cryoelectronics yielded detectors on three of three fabricated wafers. 
HYPRES delivered four of five wafers with high detector yield, with one of the original five wafers sacrificed for fabrication process development.

\section{Test Results}
\begin{figure}[!h]
\begin{center}
\includegraphics[width=\textwidth,keepaspectratio]{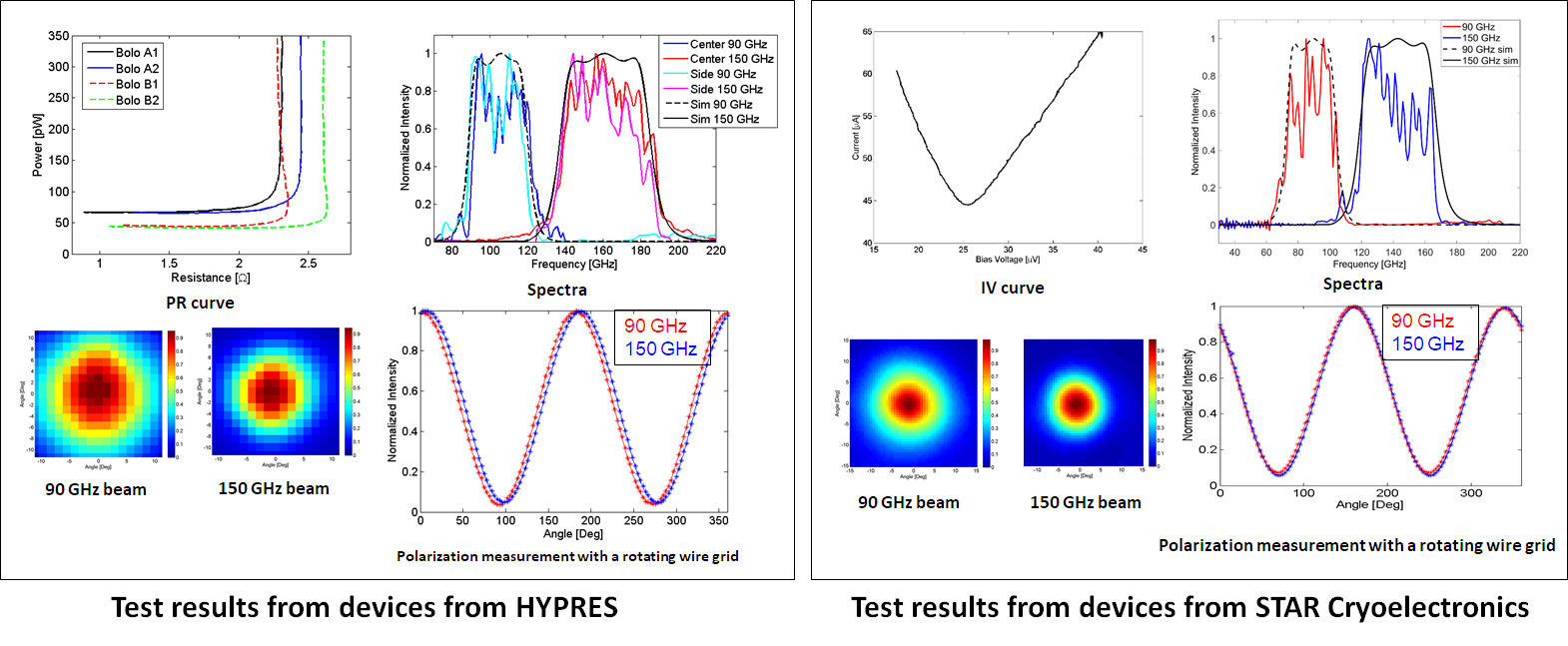}
\end{center}
\caption{(Color online) Performance of fabricated devices from HYPRES (Left) and STAR Cryoelectronics (Right). Upper left figure shows electrical bias power versus resistance curve for TES bolometers for two different frequency bands from two different locations on a wafer from HYPRES wafer. Current versus bias voltage plot is shown for STARCryoelectronics device. Upper right panel shows spectra measurement by Fourier transform spectrometer. Bottom left plot shows angular response of a multichroic detector to a temperature modulated source. Bottom right figure shows polarization response of a detector to a rotating wire grid in front of a temperature modulated source.}
\label{fig:testresult}
\end{figure}

Delivered wafers were characterized at room temperature first for DC connectivity. 
We developed a computer program for the motor actuated automatic micro probe station to probe through approximately 1,600 detectors on a detector array wafer.
We checked for thru, shorts to neighboring channel, and undesirable shorts to ground layer. 
This automated probe station can scan through 1,600 detectors in about an hour. 
We defined electrical yield as the number detectors that show nominal resistance, are
not shorted to neighbor, and do not show short to ground.
Highest electrical yield was 97\%. All wafers had yield above 90\%. 
Most yield losses were due to broken niobium traces between TES bolometers and wire bond pads. 

A brief description of the test setups are given below. A detailed description can be found elsewhere\cite{SuzukiThesis}.
Characterization of detectors at cryogenic temperature was conducted in an eight-inch Infrared Labs wet dewar with Helium-3 adsorption fridge that achieves a 250 milli-Kelvin base temperature. 
The dewar is equipped with Quantum Design DC SQUID readout electronics. 
Voltage biases to TES bolometers were provided by a battery at room temperature connected to a 20 milli-Ohm shunt resistor mounted on the 4 Kelvin stage.
For TES bolometer characterization, we enclosed detectors in 250 milli-Kelvin metal enclosure to block out stray optical power.
For optical tests, we mounted the detector chip on an anti-reflection coated synthesized elliptical alumina lens. 
The dewar has a zotefoam optical window with metal mesh low pass filters provided by Cardiff University. 
Michaelson Fourier transform spectrometer with a 254 micron thick mylar beam splitter were used to characterize spectral response.
Angular responses were characterized by scanning a 12.7 mm diameter temperature modulated source in front of a dewar window on a 2-D linear actuated stage.
Polarization responses were characterized by rotating wire grid in front of the temperature modulated source located at the peak of the antenna angular response. 

Plots of detector performances are shown in Figure~\ref{fig:testresult}.
Devices from both foundries showed the expected current to voltage responses with two distinct regions of constant resistance and constant power.
Devices from STAR Cryoelectronics were fabricated with aluminum TES that has superconducting transition temperature of approximately 1.2 Kelvin to fabricate bolometer with saturation power around 1000 pW for laboratory optical performance characterization using bright source.
Wafers from HYPRES were fabricated with an aluminum-managense film.
TES bolometers that have different weak link leg lengths (830 micron versus 630 micron) had saturation powers of 45 pW and 67 pW, respectively. This nearly follows the inverse leg length relationship.
The absolute value of the saturation power was higher than desirable because the critical temperature of TESes were higher than what we targeted. We targeted a TES critical temperature of 450 milli-Kelvin, but the tested chip had a critical temperature of 650 milli-Kelvin. 
This discrepancy was expected. 
The critical temperature of aluminum-manganese film is reported to depends on the temperature that a film is heated to during the fabrication
process, the underlaying layer's composition and sputter condition \cite{Dale}. 
Our goal was to demonstrate a working TES bolometer; we did not fine tune the heating temperature of a wafer for this round of fabrication.
It is our goal for future fabrications to fine tune heating temperature to optimize aluminum-manganese's critical temperature.
For detector array from HYPRES, we also compared uniformity across the wafer. 
TES bolometers were randomly selected from the center and very edge of detector wafer for characterization.
We saw $\pm 2\%$ variation in saturation power between two pixels; this is well within our tolerance.
TES bolometers for, two chips showed $\pm 5\%$ variation in bolometer resistance. 
This is mostly due to the wet etch process that was used to define aluminum-manganese, which is inherently difficult to control dimension to a high tolerance.
For future fabrication, we decided to adapt ion milling to define aluminum-manganese TES for higher tolerance.

Spectra measurement show that the RF filter worked as expected, as the signal from the broadband sinuous antenna is split into 90 GHz and 150 GHz frequency bands and matches well with simulation.
These filters were designed without prior measurement of material properties from thefoundries. 
Measurement of reasonable band location and shape suggests that the material properties of films from these foundries are close to what we obtain from academic foundries. 
For HYPRES wafers, we also looked at spread in band pass frequency by comparing randomly chosen pixels from the center and edge of a detector array.
Band pass center frequency shifts between center pixel and edge pixel were donward shift of 2 GHz for 90 GHz band and 1 GHz for 150 GHz band.
Bandwidths were consistent between these pixels, and band pass shapes are similar as shown in Figure~\ref{fig:testresult}.

Beam measurements show a round beam with devices from both foundries. 
Ellipticity, defined as the difference in beam width divided by the sum of beam width, were $2\pm1$\% and  $1\pm1$\% for 90 GHz and 150 GHz, respectively for devices from STAR Cryoelectroncs and $3\pm1$\% and  $2\pm1$\% for 90 GHz and 150 GHz, respectively for devices from HYPRES. 
We also looked at uniformity for HYPRES detector arrays. 
Spread in 150 GHz beam size was $4.9\pm0.2^\circ$ and $5.1\pm0.1^\circ$ for center and edge pixels, respectively.
150 GHz beam eliipticity was $3\pm1$ and $1\pm1$ for center and edge pixels, respectively.

Polarization response was characterized by rotating wire grid between the dewar and the temperature modulated source. 
Polarization leakage measurements results were setup dependent. 
Results were sensitive to how the grid was tilted and shielded respect to the dewar window. 
We report measured value as an upper limit as improved setup may show lower leakage.
For devices from STAR Cryoelectronics, we measured polarization leakage upper limits of 7\% and 6\% for 90 GHz and 150 GHz, respectively.
For devices from HYPRES, we measured upper limits of 4\% and 5\%, respectively.

\section{Conclusion and Future Developments}
We successfully designed, fabricated, and tested antenna coupled TES bolometer devices with commercial foundries. 
Transfer of fabrication processes was done successful between scientists at the national laboratory and engineers at the foundries.
Production rate was high and wafers were delivered with high yield.
Fabricated TES bolometers were successfully biased and readout by SQUID readout electronics.
Fabricated RF structures were sensitive to millimeter wave signals with their performance mostly matching with the simulations.
To push further on this approach, our next goal is to make an antenna-coupled bolometer detector array with optimized characteristics that is suitable for CMB observation. 
We would also like to demonstrate repeatability and production rate of the foundry to prepare for the next generation CMB experiment.
R\&D on fabricating superconducting devices at commercial foundries not only benefits future CMB experiment, it also provides
more fabrication options for other fields that need superconducting devicessuch as the Quantum Information Science and Dark Matter experiment. 

\begin{acknowledgements}
This work was supported by Laboratory Directed Research and Development (LDRD) funding from Berkeley Lab, provided by the Director, Office of Science, of the U.S. Department of Energy under Contract No. DE-AC02-05CH11231
We thank Dr. Daniel Yohannes, Dr. Oleg Mukhanov, and Dr. Alex Kirichenko from HYPRES Inc. for valuable suggestions and feedback that led to successful fabrication
We thank Dr. Robin Cantor from STAR Cryoelectronics for valuable suggestions and feedback that led to successful fabrication
\end{acknowledgements}


\end{document}